\documentclass{article}
\usepackage{spconf,amsmath,amssymb,graphicx,xcolor}
\usepackage{cite}
\usepackage[hidelinks]{hyperref}

\usepackage{dblfloatfix} 
\usepackage{booktabs} 
\usepackage{multirow}
\usepackage{enumitem}

\makeatletter
\let\oldthebibliography\thebibliography
\def\thebibliography#1{%
  \oldthebibliography{#1}%
  \setlength{\itemsep}{0pt}%
  \setlength{\parsep}{0pt}%
}
\makeatother
\makeatletter

\renewcommand\section{\@startsection{section}{1}{\z@}%
  {-2.0ex \@plus -1ex \@minus -.2ex}%
  {1.3ex \@plus .2ex}%
  {\normalfont\normalsize\bfseries}}
\renewcommand\subsection{\@startsection{subsection}{2}{\z@}%
  {-1.6ex \@plus -1ex \@minus -.2ex}%
  {0.8ex \@plus .2ex}%
  {\normalfont\normalsize\bfseries}}
\makeatother

\title{AV-QTSS: Class-Query Target Sound Spatialization in Multi-Source 360$^\circ$ Videos}
\name{Qingyu Luo$^1$, Peng Zhang$^1$, Davide Berghi$^2$, Wenwu Wang$^1$, Philip J.B. Jackson$^1$}
\address{$^1$Centre for Vision, Speech and Signal Processing (CVSSP), University of Surrey, U.K.\\
  $^2$L-Acoustics, U.K.}
\begin{document}
%
\maketitle
\begin{abstract}
Spatial audio is central to immersive 360$^\circ$ media, yet existing audio-visual spatialization renders the scene as a whole, leaving no control over an individual source. We introduce Audio-Visual Query-conditioned Target Sound Spatialization (AV-QTSS): given a monaural multi-source mixture, a 360$^\circ$ video and a class query from the 13-class DCASE sound-event ontology, it generates the first-order Ambisonics (FOA) signal of the queried source alone. The omnidirectional mixture contains no explicit spatial cues, so direction has to be inferred from the video, where several objects may sound at once. We propose a query-conditioned dense binding localizer that predicts a spherical probability map over the whole scene, coupled to the renderer so both are optimized jointly. Experiments with one to four concurrent sources show improved localization and spatial accuracy relative to established methods and ablated variants, and the localizer transfers to real 360$^\circ$ recordings after adaptation.

\end{abstract}

\section{Introduction}
\label{sec:intro}

Immersive media such as 360$^\circ$ video, virtual reality, and augmented reality have increased the demand for spatial audio that keeps audible events aligned with the surrounding visual scene. First-Order Ambisonics (FOA) compactly encodes a full-sphere sound field and supports flexible reproduction over loudspeakers and headphones~\cite{zotter2019ambisonics}. Whole-scene reproduction is insufficient for content creation or interactive playback, where one event may need to be isolated, enhanced, suppressed, or repositioned independently of the rest~\cite{ning2024mimosa}. Yet recording collapses all sources into a single mixture, motivating systems that recover and spatialize a user-specified sound.

\begin{figure}[t]
  \centering
  \includegraphics[width=\linewidth,]{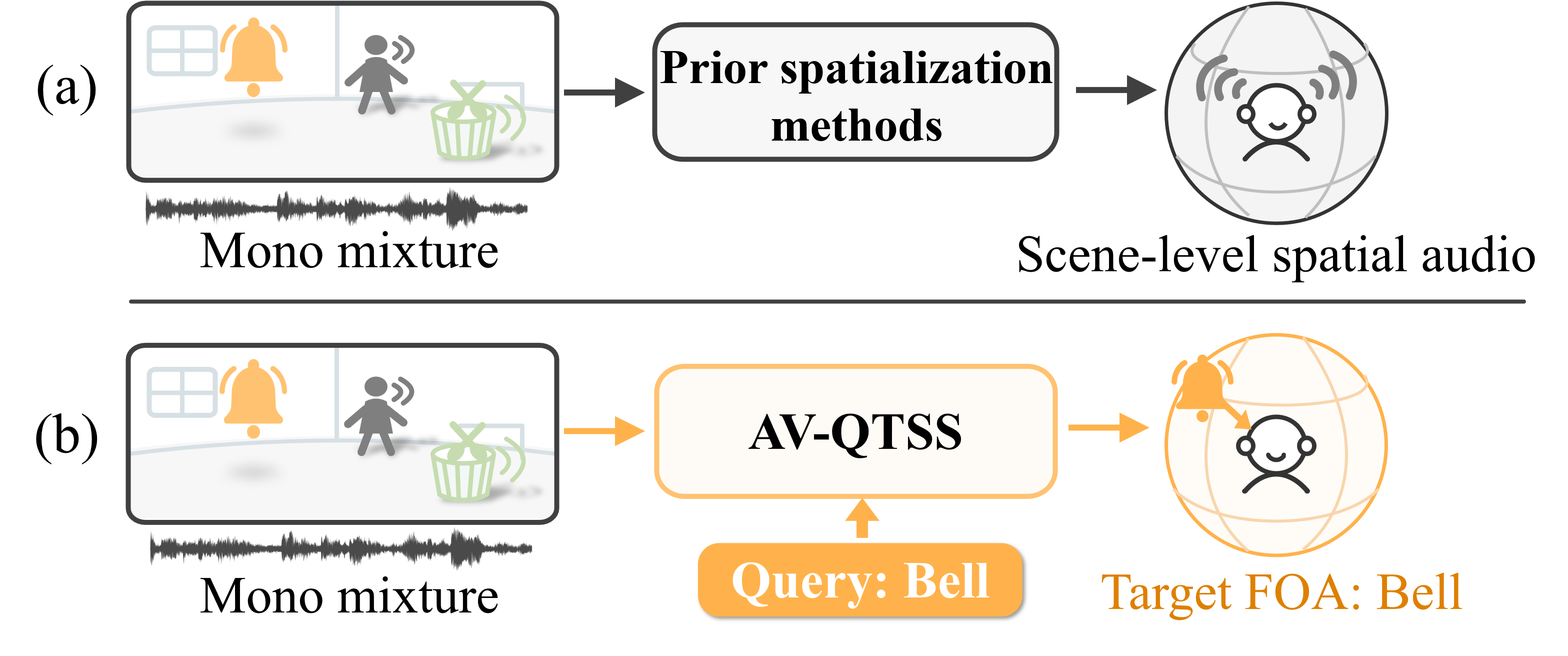}
  \vspace{-8mm}
  \caption{(a) Scene-level spatialization. (b) Class-query target-level spatialization.}
  \label{fig:fig1}
  \vspace{-1.5mm}
\end{figure}

Existing audio-visual spatialization methods usually reconstruct the whole scene from monaural audio and video. Visual cues guide the reconstruction of binaural audio for conventional video~\cite{gao20192,zhou2020sep} or FOA for panoramic content~\cite{morgado2018self,luo2026visually}. Recent video-conditioned generators such as OmniAudio and ViSAGe instead synthesize FOA soundtracks from video alone~\cite{liu2025omniaudio,kim2025visage}, producing plausible audio for the scene rather than recovering what was recorded. In both cases the spatialization objective is a scene-level soundtrack, not an independently usable output for a selected source, as contrasted in Fig.~\ref{fig:fig1}.

Source-selective spatialization requires both extracting the queried sound and locating its visual source, two tasks often addressed separately. Query-conditioned extraction isolates a target from a monaural mixture but returns it without location~\cite{delcroix2023soundbeam,audiosep}; text-queried localization returns a direction rather than a signal~\cite{zhao2024text}. Audio-visual segmentation localizes the sounding objects at pixel and instance level, but takes its audio input as given and never extracts a target from the mixture~\cite{avs,guo2025avis}.
Spatial target extraction does return a spatial target, but requires an input that is already ambisonic or binaural~\cite{chen25l_interspeech,ye2024lavss}. In multi-source scenes, an audio-conditioned localizer may still respond to competing visual sources after extraction, making consistency between the extracted sound, the bound visual source, and the rendering direction the central difficulty.

\begin{figure*}[!t]
  \centering
  \includegraphics[width=1\textwidth]{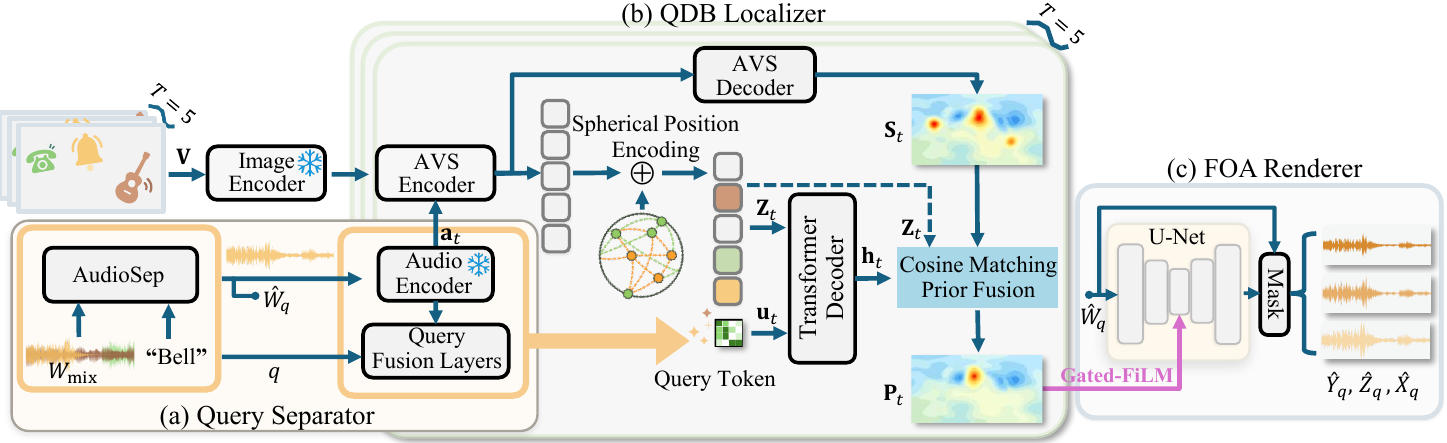}
  \vspace{-7mm}
  \caption{The framework of the proposed AV-QTSS system.}
  \label{fig:framework}
\end{figure*}

In this paper, we propose Audio-Visual Query-conditioned Target Sound Spatialization (AV-QTSS)\footnote{Demo page: \url{https://avqtss.github.io/demos/}}, in which a class query from the 13-class DCASE ontology, a monaural mixture, and a 360$^\circ$ video yield the queried source alone in FOA. Since the mixture provides limited directional information, direction has to be inferred from the video, so the same query is used both to extract the content and to bind it to the correct visual object. We therefore combine a query-conditioned Separator, a dense Localizer that predicts a query-specific spherical map, and a map-conditioned FOA Renderer.

We summarize our key contributions as follows:
\begin{itemize}[leftmargin=*,itemsep=0pt,parsep=0pt,topsep=2pt,partopsep=0pt]
\item AV-QTSS, to our knowledge the first method for class-queried target FOA generation from mono mixtures and 360$^\circ$ video, with a purpose-built 80-h corpus.
\item A Query-conditioned Dense Binding (QDB) Localizer that improves localization in multi-source scenes.
\item An evaluation showing improved localization and spatial accuracy, and real-domain transfer after adaptation.
\end{itemize}

\section{Method}
\label{sec:method}

AV-QTSS renders a single user-specified source in FOA from a monaural multi-source mixture $W_{\mathrm{mix}}$, a 360$^\circ$ video $\mathbf{V}$ and a class query $q\in\{1,\dots,13\}$. FOA represents a sound field with an omnidirectional channel $W$ and three directional channels $Y$, $Z$ and $X$. A query-conditioned Separator estimates the target channel $\hat{W}_q$ from $W_{\mathrm{mix}}$ and $q$; a Localizer turns $\hat{W}_q$, $\mathbf{V}$ and $q$ into one spherical probability map $\mathbf{P}_t$ per video frame, $t\in\{1,\dots,T\}$ with $T=5$; and a Renderer conditioned on these maps preserves $\hat{W}_q$ and predicts $\hat{Y}_q$, $\hat{Z}_q$ and $\hat{X}_q$, as shown in Fig.~\ref{fig:framework}. The task is therefore
\begin{equation}
\mathcal{F}:\;\left(W_{\mathrm{mix}},\,\mathbf{V},\,q\right)
\;\mapsto\;\hat{\mathbf{F}}_q
=\left[\hat{W}_q,\,\hat{Y}_q,\,\hat{Z}_q,\,\hat{X}_q\right].
\end{equation}

\subsection{Query-Conditioned Separator}
\label{ssec:separator}

The number of active sources is unknown, so this stage must extract the content of a user-specified class rather than separate a fixed number of sources. We adopt AudioSep~\cite{audiosep}, a text-queried ResUNet separator that fits a class-query interface directly.

Two conditions differ from its original setting: our query vocabulary is fixed rather than free-form, and the target is the reverberant omnidirectional channel of the queried source, which the released model was not trained to recover. We therefore precompute the class conditions with a frozen CLAP~\cite{clap} text encoder, giving prototypes $\mathbf{C}_{\mathrm{base}}\in\mathbb{R}^{13\times512}$, and learn a zero-initialized residual $\Delta\mathbf{C}$ of the same size:
\begin{equation}
\mathbf{C}[q]=\frac{\mathbf{C}_{\mathrm{base}}[q]+\Delta\mathbf{C}[q]}
{\lVert \mathbf{C}_{\mathrm{base}}[q]+\Delta\mathbf{C}[q]\rVert_2}.
\end{equation}
The condition $\mathbf{C}[q]$ is then retrieved by index at inference.

\subsection{Query-Conditioned Dense Binding Localizer}
\label{ssec:localizer}
This stage must decide which spatial region belongs to the queried source. The AVS model~\cite{avs}, which we run on 2:1 equirectangular projection (ERP) frames with frozen PVT-v2 image and VGGish audio encoders,  already localizes sounding regions from $\hat{W}_q$, but its output sounding map $\mathbf{S}_t$ carries no explicit class identity and can remain ambiguous when several visible objects sound at once.

We therefore introduce the Query-conditioned Dense Binding (QDB) Localizer, which predicts a distribution over the whole grid. The audio-visual features of the AVS encoder are pooled at each frame to a $7\times14$ spherical grid whose $98$ cells are indexed by $j$, and projected to one 128-dimensional token per cell encoding its correspondence with the separated audio; adding a linear embedding of each cell's fixed spherical unit direction $\mathbf{g}_j\in\mathbb{R}^{3}$ gives spatially aware tokens $\mathbf{Z}_t\in\mathbb{R}^{98\times128}$.
The query fusion layers concatenate the separated-audio feature $\mathbf{a}_t\in\mathbb{R}^{128}$ of the $t$-th second and a learned class embedding $\mathbf{e}_q\in\mathbb{R}^{64}$, layer-normalize (LN) and project them into a single query token $\mathbf{u}_t\in\mathbb{R}^{128}$, so that the target is specified by its sound and its class at once:
\begin{equation}
\mathbf{u}_t=\mathrm{Linear}\!\left(\mathrm{LN}\!\left(\left[\mathbf{a}_t;\,\mathbf{e}_q\right]\right)\right).
\end{equation}
A Transformer decoder lets $\mathbf{u}_t$ read $\mathbf{Z}_t$, compressing the scene into a single target vector $\mathbf{h}_t\in\mathbb{R}^{128}$; its cosine similarity with each cell token $\mathbf{z}_{t,j}\in\mathbb{R}^{128}$ projects it back onto the grid, and the resulting scores are fused with $S_{t,j}$, the AVS prior at the same cell, in the logit domain:
\begin{equation}
P_{t,j}=\mathrm{Softmax}_j\!\left(
\alpha_s\cos\!\left(\mathbf{h}_t,\,\mathbf{z}_{t,j}\right)
+\alpha_p\log\!\left(S_{t,j}+\epsilon\right)\right).
\end{equation}
$P_{t,j}$ is the probability that the queried source occupies cell $j$. The learned scales $\alpha_s,\alpha_p>0$ set the balance of the two terms and the sharpness of $\mathbf{P}_t$. The terms are complementary: the prior drives down silent cells, while the query score separates the queried source from the other sounding ones, so the fused map favors cells supported by both.

\subsection{FOA Renderer}
\label{ssec:renderer}
Our scene-level Renderer~\cite{luo2026visually} spatializes the entire mixture: a TF U-Net predicts one complex mask per directional channel from the STFT of $W$, conditioned through gated FiLM on a class-agnostic AVS prior. We adapt it to query-specific conditioning: the dense map $\mathbf{P}_t$ of the queried source, linearly interpolated to the STFT frames, replaces this prior, and the gate follows the peak probability and entropy of the map. Applied to the STFT of $\hat{W}_q$, the masks let each channel inherit its time--frequency structure while encoding the source direction, and an inverse STFT gives $(\hat{Y}_q,\hat{Z}_q,\hat{X}_q)$.

The Localizer and Renderer are optimized jointly, so the FOA loss shapes the map via this differentiable condition:
\begin{gather}
\label{eq:loss}
\mathcal{L}_{\mathrm{joint}}=\mathcal{L}_{\mathrm{loc}}+\lambda_{\mathrm{FOA}}\mathcal{L}_{\mathrm{FOA}},\\
\label{eq:loc}
\mathcal{L}_{\mathrm{loc}}=\mathcal{L}_{\mathrm{map}}+\lambda_{\mathrm{dir}}\mathcal{L}_{\mathrm{dir}}+\lambda_{\mathrm{dist}}\mathcal{L}_{\mathrm{dist}},
\end{gather}
where $\mathcal{L}_{\mathrm{map}}$ is a cross-entropy against a normalized spherical Gaussian with $\sigma=10^{\circ}$ centered on the target,
$\mathcal{L}_{\mathrm{dir}}$ is a cosine loss on the expected direction, and $\mathcal{L}_{\mathrm{dist}}$ penalizes probability weighted by the pointwise maximum of $\sigma=20^{\circ}$ Gaussians centered on other active sources; $\mathcal{L}_{\mathrm{FOA}}$ is an $\ell_1$ loss on the real, imaginary and $0.1$-weighted magnitude STFTs of the directional channels.

\section{Experiment}
\label{sec:Experiment}

\subsection{Datasets}
\label{ssec:data}
We use a synthetic corpus for per-source supervision and real recordings for localization transfer:

\textbf{STAIRS26.}
Released 360$^\circ$ spatial-audio corpora carry scene-level ambisonics without per-source signals~\cite{liu2025omniaudio,kim2025visage}. STAIRS26~\cite{roman2026stairs26}, which extends STARSS23~\cite{shimada2023starss23}, comes closest to what this task requires: 7.5 h of real recordings with up to four concurrent sources, annotated with the 13 DCASE sound-event classes (e.g., female speech, telephone, footsteps, and musical instrument) and their directions.
We keep the 156 recordings with 360$^\circ$ video under a room-disjoint 94/27/35 split.
Source-wise waveforms and FOA are unavailable, so it supervises the one stage its annotations reach: real-domain query localization.

\textbf{AV-QTSS-Syn.}
To obtain source-wise supervision, we construct an 80-h corpus of 5-s clips under the same ontology using SpatialScaper~\cite{roman2024spatial} and SELDVisualSynth~\cite{roman2025generating}.
Clips provide a monaural mixture and reverberant per-source targets. 
The audio-only portion provides varied overlaps for Separator adaptation; the AV-complete portion adds 360$^\circ$ video, source-wise FOA and trajectories, with each source present throughout for framewise spatial supervision.
Each 40-h portion contains 24,000/2,400/2,400 train/validation/test clips.
We curate 3,904 audio recordings, 3,804 visual-object identities, and 1,984 indoor panoramas from a broad public-resource pool, led by FSD50K, EARS and FMA~\cite{fonseca2021fsd50k,richter2024ears,defferrard2016fma,mei2024wavcaps,piczak2015esc} for audio, Open Images and Flickr30k~\cite{kuznetsova2020openimage,plummer2015flickr30k,gupta2019lvis} for visual objects, and 360-Indoor~\cite{chou2020360indoor} for panoramas.
All assets are filtered, deduplicated and reviewed; foreground tiles are softly blended into panoramas to reduce seams~\cite{berghi2025stereo}. We collect and validate 20,627 direction-dependent FOA RIR grid entries from 26 measured-room variants, plus 105 native-FOA ambience recordings, led by DCASE 2026 Task 4, TAU-SRIR/TAU-SNoise and SurrRoom~\cite{dcase2026t4,tau_srir,cieciura2023surrroom,mckenzie2021dataset,green2017eigenscape}. Concurrent sources $K{=}1/2/3/4$ occur in 10/40/35/15\% of clips; motion is split 50/50\% static/moving for $K{=}1$ and 25/50/25\% all-static/mixed/all-moving for $K{\ge}2$; 50\% include ambience at 5--20 dB SNR.
All resource identities are disjoint across splits, and all synthetic results use the AV-complete test split.

\subsection{Experimental Setup}
\label{ssec:Setup}

We use 5-s, 24-kHz audio and five synchronized ERP frames sampled at 1 Hz and resized to $224\times448$. FOA follows the ACN/SN3D convention with channel order $[W,Y,Z,X]$. The Localizer has a two-layer Transformer decoder with four attention heads. On synthetic data, we train the query-conditioned Separator independently through staged fine-tuning with module-specific learning rates from $2.5\times10^{-7}$ to $2.5\times10^{-5}$. We then jointly optimize the Localizer and Renderer on target waveforms estimated by the Separator, using AdamW with cosine decay, a learning rate of $2\times10^{-5}$, weight decay of $10^{-4}$, gradient clipping at 1.0, and $\lambda_{\mathrm{dir}}=1$, $\lambda_{\mathrm{dist}}=1.75$, $\lambda_{\mathrm{FOA}}=20$. For STAIRS26, we update only the QDB Localizer, at $2\times10^{-5}$.

\subsection{Results}
\label{ssec:Results}

Reconstruction metrics are waveform MSE $\ell_2$, magnitude and wrapped-phase RMSE $\mathcal{L}_{\mathrm{mag}}$ and $\mathcal{L}_{\mathrm{phs}}$, and multi-resolution STFT distance $\mathcal{L}_{\mathrm{MRS}}$. All are computed on the directional channels. For $W$, SI-SDR against the reverberant target is $-3.75$ dB for the mixture and $6.33$ dB after the Separator on multi-source clips. Spatial metrics use DOA estimation~\cite{liu2025omniaudio,luo2026visually}, reporting mean absolute azimuth and elevation errors $\Delta_{\mathrm{abs}}\theta$ and $\Delta_{\mathrm{abs}}\phi$, and overall angular error $\Delta_{\mathrm{ang}}$. MOS-C and MOS-S assess content correctness and perceived spatial accuracy: 11 participants rate five binaurally decoded clips per method on a 1--5 scale, averaged over listeners and clips.

\begin{table*}[t]
  \centering
  \caption{Comparison of Separator and Renderer methods, and ablation of Localizer components, on AV-QTSS-Syn. Except for the training-free Ambi-Encode, all variants are retrained under the same schedule. Best in \textbf{bold}, second best \underline{underlined}.}
  \label{tab:system}
  \small
  \setlength{\tabcolsep}{6pt}
  \begin{tabular}{l cccc ccc cc}
    \toprule
    \multicolumn{1}{c}{\textbf{Model}} &
    \multicolumn{4}{c}{\textbf{Reconstruction Metrics}} &
    \multicolumn{3}{c}{\textbf{Spatial Metrics}} &
    \multicolumn{2}{c}{\textbf{Subjective Metrics}}\\
    \cmidrule(lr){2-5}\cmidrule(lr){6-8}\cmidrule(lr){9-10}
     & $\boldsymbol{\ell_2(\times 10^{3})\downarrow}$
     & $\boldsymbol{\mathcal{L}_{\mathrm{mag}}\downarrow}$
     & $\boldsymbol{\mathcal{L}_{\mathrm{phs}}\downarrow}$
     & $\boldsymbol{\mathcal{L}_{\mathrm{MRS}}\downarrow}$
     & $\boldsymbol{\Delta_{\mathrm{abs}}\theta\downarrow}$
     & $\boldsymbol{\Delta_{\mathrm{abs}}\phi\downarrow}$
     & $\boldsymbol{\Delta_{\mathrm{ang}}\downarrow}$
     & \textbf{MOS-C} $\boldsymbol{\uparrow}$
     & \textbf{MOS-S} $\boldsymbol{\uparrow}$ \\
    \midrule
    \multicolumn{10}{c}{\textbf{Comparison to Separator Methods}}\\[0.5ex]
    No-Separator        & 21.7 & 1.50 & 1.75 & 4.70 & 29.5° & 19.1° & 39.2° & 2.33 & 2.95 \\
    USS~\cite{kong2023uss}    & 3.66 & 0.82 & 1.71 & 3.43 &	19.4° & 12.4° & 26.4° & 3.05 & 3.36 \\
    \midrule  
    \multicolumn{10}{c}{\textbf{Ablation on Localizer Components}}\\[0.6ex]
    No-Binding~\cite{avs}    & 4.07 & 0.62 & 1.70 & 3.20 & 43.9° & 13.5° & 47.4° & 3.85 & 2.44 \\
    No-Class      & 3.83 & 0.59 & 1.67 & 3.18 & 34.3° & 13.2° & 38.7° & 4.04 & 2.80 \\
    \midrule
    \multicolumn{10}{c}{\textbf{Comparison to Renderer Methods}}\\[0.6ex]
    Ambi-Encode~\cite{zotter2019ambisonics}     & 3.53 & 0.60 & \textbf{1.63} & 3.50 & \textbf{17.3°} & 12.6° & 25.5° & 3.75 & 3.73 \\
    Direction-Only      & \underline{3.21} & \textbf{0.57} & \underline{1.64} & 3.24 & 18.4° & \underline{12.1°} & \underline{24.4°} & 3.84 & \underline{3.91} \\
    \midrule
    Ours (No-Joint) & 3.49 & 0.61 & 1.66 & \underline{3.12} & 23.4° & \textbf{11.5°} & 27.7° & \underline{4.09}   & 3.87 \\
    \textbf{Ours}  & \textbf{3.01} & \underline{0.58}  & \textbf{1.63} & \textbf{3.11} & \underline{17.4°} & 12.3° & \textbf{23.3°} & \textbf{4.24} &  \textbf{4.15} \\
    \bottomrule
  \end{tabular}
\end{table*}

As no prior system performs the complete task, Table~\ref{tab:system} evaluates the full system with one stage replaced or ablated at a time. No-Separator uses the mixture; USS~\cite{kong2023uss} uses the same class-vector condition and is adapted on our corpus. The Localizer ablations assess QDB refinement and explicit class conditioning: No-Binding cascades the existing components, replacing the QDB map with the AVS prior~\cite{avs} from the separated target audio; No-Class removes only the Localizer's class embedding.
Ambi-Encode analytically encodes $\hat{W}_q$ at the expected direction of the map~\cite{zotter2019ambisonics}; Direction-Only retains only that direction, which stays differentiable; Ours (No-Joint) trains the Localizer and Renderer separately.
Without separation, interference raises $\ell_2$ to 21.7 and worsens $\mathcal{L}_{\mathrm{mag}}$ and $\mathcal{L}_{\mathrm{MRS}}$. USS raises $\mathcal{L}_{\mathrm{mag}}$ to 0.82 and lowers MOS-C to 3.05. No-Binding and No-Class raise $\Delta_{\mathrm{ang}}$ from 23.3$^{\circ}$ to 47.4$^{\circ}$ and 38.7$^{\circ}$, with larger drops in MOS-S than MOS-C. Joint training reduces $\Delta_{\mathrm{ang}}$ by 4.4$^{\circ}$ versus No-Joint. Ours achieves the lowest overall angular error and the strongest overall balance across the three groups.
 
\begin{table}[t]
  \centering
    \caption{QDB Localizer ablation on AV-QTSS-Syn and on real STAIRS26 recordings after adaptation, by number of concurrent sources $K$ (Acc@30$^{\circ}$ \%$\uparrow$).}
  \label{tab:stage}
  \small
  \setlength{\tabcolsep}{4pt}
  \renewcommand{\arraystretch}{1.15}
  \begin{tabular*}{\columnwidth}{@{\extracolsep{\fill}}llcccc}
    \toprule
    \textbf{Dataset} & \textbf{Localizer} & $K{=}1$ & $K{=}2$ & $K{=}3$ & $K{=}4$ \\
    \midrule
    \multirow{3}{*}{AV-QTSS-Syn}
      & No-Binding~\cite{avs}   & 97.0 & 76.6 & 65.5 & 55.1 \\
      & No-Class     & 97.1 & 78.5 & 69.9 & 60.4 \\
      & \textbf{QDB} & \textbf{97.9} & \textbf{90.6} & \textbf{89.6} & \textbf{82.7} \\
    \midrule
    \multirow{3}{*}{STAIRS26}
      & No-Binding~\cite{avs}    & 38.8 & 44.0 & 47.1 & 19.7 \\
       & No-Class                & 42.3 & 58.1 & 57.9 & 20.1 \\
      & \textbf{QDB} & \textbf{55.0} & \textbf{69.4} & \textbf{60.2} & \textbf{29.6} \\
    \bottomrule
  \end{tabular*}
\end{table}
                   
Table~\ref{tab:stage} evaluates the Localizer alone. Acc@30$^{\circ}$ is the fraction of query frames whose map peak lies within $30^{\circ}$ of the annotation. Synthetic mixtures use distinct classes; on STAIRS26, frames with several active instances of the queried class are excluded. Lacking source-wise references, STAIRS26 supervises localization only, and each Localizer variant is adapted under the same schedule.
On synthetic data, No-Class improves on No-Binding, while adding the Localizer's class embedding yields gains that grow with $K$. The same ranking holds after adaptation on STAIRS26, where absolute accuracy remains low at $K{=}4$.
Without adaptation the QDB Localizer stays between 11.1 and 32.8\%, so adaptation at least doubles Acc@30$^{\circ}$ in every subset, where sounding objects are small and often not visually salient.
On STAIRS26, accuracy is not monotonic in $K$ for any variant: unlike the synthetic subsets, the real ones differ in scene and class content, not only in polyphony.

\begin{figure}[t]
  \centering
  \includegraphics[width=\linewidth]{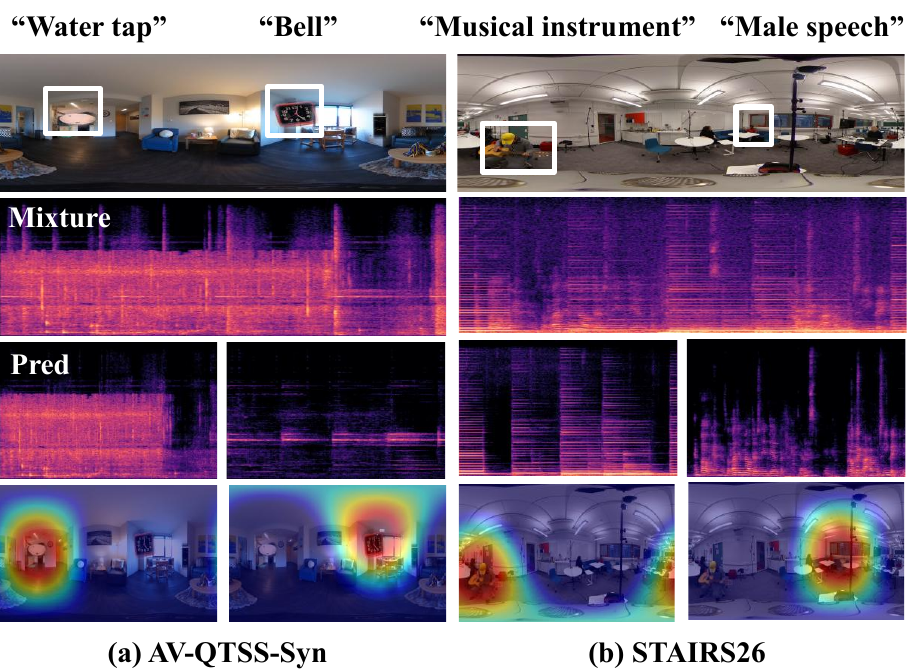}
  \vspace{-8mm}
 \caption{Qualitative results. Rows: input scene, mixture, predicted target, rendered FOA energy map. Boxes mark the annotated position of each queried source.}
  \label{fig:fig3}
\end{figure}

Fig.~\ref{fig:fig3} shows query switching in three-source synthetic (a) and two-source real (b) scenes. With each scene and mixture fixed, the extracted content and rendered energy follow the queried source on both datasets, consistent with the localization results. SI-SDR in Fig.~\ref{fig:fig3}(a) is 9.6\,dB for the water tap (third row left) and 6.2\,dB for the bell (right).

\section{Conclusion}
\label{sec:Conclusion}

We presented AV-QTSS for rendering a user-specified source in FOA from a mono mixture and 360$^\circ$ video, with an 80-h corpus for per-source supervision and a QDB Localizer coupled to the Renderer for joint optimization. Experiments show improved localization and spatial accuracy of the rendered FOA relative to established methods and ablated variants, and real-domain transfer after adaptation.
The formulation assumes closed-set class queries that are active in the clip, cannot separate same-class instances, and may confuse semantically close classes. Future work will target open-vocabulary queries, higher polyphony, and realistic corpora.

\section{Acknowledgment}
\label{sec:acknowledgment}

We thank Pablo Martínez-Nuevo, Sven Ewan Shepstone, and Jon Francombe (Bang \& Olufsen A/S) for valuable discussions. This research was supported by Bang \& Olufsen A/S as part of the AURIC project.




{\small
\bibliographystyle{IEEEbib}
\bibliography{strings,refs}
}

\end{document}